\documentstyle[preprint,aps]{revtex}
\def\be{\begin{equation}}
\def\ee{\end{equation}}
\def\bea{\begin{eqnarray}}
\def\eea{\end{eqnarray}}

\begin{document}
\draft

\preprint{IP/BBSR/95-45}
\title {\bf S-DUALITY AND THE ENTROPY OF BLACK HOLES IN HETEROTIC
STRING THEORY }
\author{ Amit Ghosh$^a$\footnote{e-mail: amit@tnp.saha.ernet.in} and
Jnanadeva Maharana$^b$\footnote{e-mail: maharana@iopb.ernet.in}}
\address{$^a$Saha Institute of Nuclear Physics, 1/AF Bidhan Nagar,
Calcutta 700 064, INDIA.\\
$^b$Institute of Physics, Bhubaneswae 751005,
INDIA.}
\maketitle
\begin{center}
April 30, 1995
\end{center}
\begin{abstract}
Four dimensional heterotic string effective action is
known to admit
non-rotating electrically and magnetically charged black hole
solutions. It is shown that the partition functions and
entropies for both the cases are identical when these black hole
solutions are related by S-duality transformations.
The entropy is computed
and is vanishing for each black hole in the extremal limit.

\end{abstract}
\vspace{2 cm}

\narrowtext

\newpage


\par Recently, Hawking and Ross [1] have implemented S-duality
transformation to derive interesting results for the
electrically and magnetically charged black holes. It is well
known that the partition function for electrically charged black
holes is analogous to that of the grand canonical ensemble with
a chemical potential; whereas the partition function for the
magnetically charged black holes has the interpretation of
canonical partition function. Naturally, the two partition
functions differ from one another. In order to investigate the
implications of S-duality the charged black hole partition
function is projected to one with definite charge and then it is
found that the partition functions for the electrically and
magnetically charged black holes are identical.

\par It is recognized that the consequences of S-duality are
interesting and surprising [2]. In the recent past, considerable
attention has been focussed to investigate implications of
S-duality in sring theory [3] and supersymmetric field theories
[4].
This transformation, in its simplest form, when applied to
electrodynamics, interchanges the roles of electric and magnetic
fields. However, in theories with larger number of field
contents, the transformation rules for the fields are to be
appropriately defined so that the equations of motion
remain invariant under S-duality transformations. Note however,
that the action is not necessarily invariant under these
transformtions.  If we
toroidally compactify ten dimensional heterotic string effective
action [5] to four dimensions, the reduced effective action is
endowed with N=4 supersymmetry. One of the attractive features
of the four dimensional theory is that the quantum corrections
are restricted due to the existence of non-renormalisation
theorems. It is expected that this theory respects
strong-weak coupling duality symmetry and might provide deeper
insight into the non-perturbative aspects of string theory.
Furthermore, the four dimensional reduced action can be cast in
a manifestly O(6,22) invariant form. Cosequently, one can
generate new background field configurations by implementing
suitable O(6,22) and S-duality tranformations on a known
solution of the effective action. String effective action admit
black hole solutions[6].
Sen [7] has obtained interesting solutions such that they
preserve N=4 supersymmetry, saturating Bogomol'nyi
bound in the extremal limit.

\par The purpose of this letter is to investigate the
consequences of S-duality transformations for the black hole
solutions of the four dimensional heterotic string effective
action and investigate the thermodynamic properties of these
black holes. First, we compute the partition function of an
electrically charged black hole which resembles a grand
partition function with a chemical  potential. Then we adopt the
techniques due to Coleman Preskill and Wilczek [8] to project to
a partition function with definite charges. The free energy,
entropy and
the thermodynamical potential are derived. The entropy is shown
to be zero in the extremal limit. Our result is of importance due to
the fact that these extremal black hole solutions saturate
Bogomol'nyi bound and the partition functions are expected not
to get quantum corrections due to N=4 supersymmetry. Our next
result provide yet another proof of S-duality. We envisage
electrically charged black hole solutions with zero axion field.
Subsequently we implemented the S-duality tranformation such
that the electrically charged solutions go over to the magnetic
ones. Furthermore, the charged projected partition function for
electrically charged black hole is identical to that of the
magnetically charged black hole. While calculating the partition
functions we use the Euclidean formulation throughout the paper.

The massless bosonic sector of four dimensional heterotic
string effective action consist of
dialton, graviton, axion (dual to the antisymmetric tensor
field) and 28 abelian gauge fields denoted respectively by $\phi$,
$G_{\mu\nu}$, $\psi$ and $A_\mu^{(a)}$, $a=1\cdots 28$. The
effective action, written in Einstein metric, $g_{\mu\nu}=e^
{-\phi} G_{\mu\nu}$ takes the following form,

\bea  S&= &\int d^4x \sqrt {-g}
\bigg\{ R + {1\over 2\lambda_2^2}
\partial_{\mu}\bar\lambda\partial^{\mu}\lambda
+ {1 \over 8} {\rm tr} (\partial_\mu ML \partial^
\mu
ML) \nonumber\\&-&\lambda_2
F^{(a)}_{\mu \nu} (LML)_{ab} F^{(b)\mu
\nu}+\lambda_1F_{\mu\nu}^{(a)} L_{ab}\tilde F^{(b)\mu\nu} \bigg\}
+~~{\rm Boundary~~ terms} \eea

Where,

\bea  F^{(a)}_{\mu \nu} = \partial_\mu A^{(a)}_\nu - \partial_\nu
A^{(a)}_\mu. \eea

\noindent $M$ and $L$ are 28$\times$28 symmetric matrices; whereas, $M$
paramtrizes the coset ${O(6,22)\over O(6)\otimes O(22)}$, $L$
has 22 eigenvalues $-1$ and 6 eigenvalues $+1$ and  has the
following form

\bea L=\left (\matrix {-I_{22} & 0\cr
0 & I_6\cr}\right ).\eea

\noindent $M$ satisfies the property, $MLM^T=L$.  $\lambda$ is a
complex scalar field defined in terms of the axion and the dilaton
$\lambda=\psi + ie^{-\phi}$. The equations of motion associated
with (1) remain invariant under the S-duality transformations

\bea\lambda &\to &\lambda'={a\lambda+b\over c\lambda+d},\qquad
ad-bc=1\nonumber\\
F_{\mu\nu}^{(a)}&\to &(c\lambda_1+d)F_{\mu\nu}^{(a)} +
c\lambda_2(ML)_{ab} \tilde F_{\mu\nu}^{(a)}.\eea

where a,b,c and d are real. The backgrounds $g_{\mu\nu}$ and M
remain invariant under the transformation.

The invariance property of the action under the non-compact
global transformations has been utilised by Sen to generate
charged rotating black hole solutions
[7] starting from an uncharged rotating black hole solution
with $\phi=0$, $B_{\mu\nu}=0$, $A_\mu^{(a)}=0$, $M=I_{28}$.
A special case corresponding to non-rotating
charged black hole is given by the following
background configurations,

\bea ds^2=-\Delta^{-1/2}(r^2-2mr)dt^2+\Delta^{1/2}(r^2-
2mr)^{-1}dr^2+\Delta^{1/2}d\Omega_{II}^2\eea
where,
$\Delta=r^4+2mr^3(\cosh\alpha\cosh\gamma-1)+m^2r^2
(\cosh\alpha-\cosh\gamma)^2$.
The dilaton is given by,
\bea e^{\phi}=r^2/\Delta^{1/2}\eea
The gauge fields are,

\bea A_t^{(a)}&=&-{n^{(a)}\over\sqrt
2}{mr\over\Delta}\sinh\alpha[r^2\cosh\alpha+mr(\cosh\alpha
-\cosh\gamma)]\qquad 1\le a\le 22\nonumber\\ &=&-{p^{(a-22)}\over
\sqrt2}{mr\over\Delta}\sinh\gamma[r^2\cosh\alpha+mr(
\cosh\gamma-\cosh\alpha)]\qquad 23\le a\le 28\eea
The M-matrix, satisfying the equations of motion is,
\bea M=I_{28}+\left (\matrix {Pnn^T & Qnp^T\cr
Ppn^T & Ppp^T\cr}\right )\eea
where, $P=2{m^2r^2\over\Delta}\sinh^2\alpha\sinh^2\gamma$ and
$Q=2{mr\over\Delta}\sinh\alpha\sinh\gamma[r^2+mr
(\cosh\alpha\cosh\gamma -1)]$.
The mass of this black hole is
\bea M={m\over
2}(1+\cosh\alpha\cosh\gamma)\eea
The 28 charges are given by

\bea Q^{(a)}&=&{n^{(a)}\over\sqrt2}m\sinh\alpha\cosh\gamma\qquad
1\le a\le 22\nonumber\\ &=&{p^{(a-22)}\over
\sqrt2}m\sinh\gamma\cosh\alpha\qquad 23\le a\le 28\eea

Note that the ``boost'' angles $\alpha,\gamma$ are the parameters
[7,9] that appear in global non-compact transformation to obtain charged
solution from the uncharged one.

The horizon of this black hole is at $r=2m$, there is a
curvature singularity at $r=0$, the area of the event horizon is
$A_H=8\pi m^2(\cosh\alpha+\cosh\gamma)$ and the
Hawking-temperature is given by $T_H={1\over 4\pi
m(\cosh\alpha+\cosh\gamma)}$.

There are two extremal limits of this black hole which preserve
supersymmetry and consequently saturate the Bogomol'nyi bound for the
mass:

(I) $m\to 0$, $\gamma\to\infty$, while keeping $m\cosh\gamma=m_0$
finite and $\alpha$ is finite but arbitrary.
For this case $\Delta=r^2(r^2+2m_0r\cosh\alpha+m_0^2)$, the mass
is $M={m_0\over 2}\cosh\alpha$ and the charges are

\bea Q_L^{(a)}={n^{(a)}\over\sqrt2}m_0\sinh\alpha\quad
1\le a\le 22,\qquad Q_R^{(a)}={p^{(a-22)}\over\sqrt2}m_0
\cosh\alpha\quad 23\le a\le 28\eea
The horizon of this black hole is at $r=0$, hence $A_H=0$ and
temperature , $T_H={1\over 8\pi M}\cosh\alpha$.

The expression for the gauge fields become,

\bea A_{tL}^{(a)}=-{n^{(a)}\over\sqrt
2}{mr^3\over\Delta}\sinh\alpha
\quad 1\le a\le 22,\quad A_{tR}^{(a)}=-{p^{(a-22)}\over
\sqrt2}{mr^3\over\Delta}\cosh\alpha\quad 23\le a\le 28\eea

The Bogomol'nyi bound is saturated and $M^2={1\over
2}\vec Q_R^2$, where R stands for the right hand sector and
the index $a$ runs from 23 to 28. Also $Q_L^{(a)}=\sqrt 2
Mn^{(a)}\tanh\alpha$.

(II) The other black hole corresponds to limits: $m\to 0$,
$\alpha= \gamma
\to\infty$ such that $m\cosh^2\alpha=m_0$ is finite.
The parameters are $\Delta=r^2(r^2+2m_0r)$,~ $M={m_0\over
2}$,

the cherges are,

\bea Q_L^{(a)}={n^{(a)}\over\sqrt2}m_0\quad
1\le a\le 22,\qquad Q_R^{(a)}={p^{(a-22)}\over\sqrt2}m_0
\quad 23\le a\le 28.\eea
The horizon is at $r=0$, so $A_H=0$ and
$T_H=\infty$. The gauge fields are given by,

\bea A_t{(a)}&=&-{n^{(a)}\over\sqrt
2}{mr^3\over\Delta}
\qquad 1\le a\le 22\nonumber\\ &=&-{p^{(a-22)}\over
\sqrt2}{mr^3\over\Delta}\qquad 23\le a\le 28\eea
The mass saturates the Bogomol'nyi bound $M^2={1\over 2}\vec
Q_R^2={1\over 2}\vec Q_L^2$.

We mention in passing that the thermodynamic properties
follow naturally from
the non-extremal solution. Define $\Phi^{(a)}=A_t^{(a)}|_{r=2m}$
to be the elctrostatic potential at the horizon. Then the
following  relations hold.

\bea T_Hd{A_H\over 4}&=&dM-\sum_a\Phi^{(a)}dQ^{(a)} \nonumber\\
 T_H{A_H\over 2}&=&M-\sum_a\Phi^{(a)}Q^{(a)}\eea
The explicit form of $\Phi^{(a)}$ is given by,

\bea \Phi_L^{(a)}&=&{n^{(a)}\over\sqrt
2}{\sinh\alpha\over \cosh\alpha+\cosh\gamma}
\qquad 1\le a\le 22\nonumber\\ \Phi_R^{(a)}&=&{p^{(a-22)}\over
\sqrt2}{\sinh\gamma\over\cosh\alpha+\cosh\gamma}
\qquad 23\le a\le 28\eea

For case (I),

\bea \Phi_L^{(a)}=0\quad 1\le a\le 22,\qquad \Phi_R^{(a)}=
{p^{(a-22)}\over\sqrt2}
\quad 23\le a\le 28\eea

and for case (II),

\bea \Phi_L^{(a)}={n^{(a)}\over 2\sqrt2}
\quad 1\le a\le 22,\qquad \Phi_R^{(a)}={p^{(a-22)}\over 2
\sqrt2}\quad 23\le a\le 28\eea
Now we turn to equations of motion,

\bea R_{\mu\nu}={1\over
2\lambda_2^2}(\partial_{\mu}\bar\lambda\partial_{\nu}
\lambda+ \partial_{\nu}\bar\lambda\partial_{\mu}\lambda)&+&
{1\over 8}Tr(\partial_{\mu}ML\partial_{\nu}ML)-g_{\mu\nu}
{1\over 16}Tr(\partial_{\mu}ML\partial^{\mu}ML)\nonumber\\
&+&2\lambda_2F_{\mu\rho}^{(a)}(LML)_{ab}F_{\nu}^{(b)\rho}-{1\over 2}
\lambda_2g_{\mu\nu}F_{\rho\sigma}^{(a)}(LML)_{ab}F^{(b)\rho\sigma}
\nonumber\\D_{\mu}(-\lambda_2(ML)_{ab}
F^{(b)\mu\nu}&+&\lambda_1\tilde
F^{(a)\mu\nu} )=0\eea
where $\tilde F^{(a)\mu\nu}={1\over 2\sqrt{-g}}\epsilon^{\mu\nu
\rho\sigma}F^{(a)}_{\rho\sigma}$ is the usual dual tensor
satisfying Bianchi identities.

The first equation in (19) corresponds to the Einstein equation,
where
 $R$ has been eliminated in favour of the matter energy-momentum stress
tensor and the second one  corresponds to the gauge field equation.

In order to compute the partition function, we need to determine
the action on shell. One of the efficient ways to obtain this
action is to take the trace of $R_{\mu \nu}$ in (19) to obtain
the scalar curvature, R,
and use the expression in (1). A straightforward calculation
shows that the contributions comes from the well known
gravitational boundary term as well as  from the gauge field surface term

\bea {1\over 8\pi}\int_{\Sigma^{\infty}}\sqrt{-g} d^3x~
n_r[-\lambda_2A_t^{(a)}
(LML)_{ab} F^{(b)tr}+\lambda_1
A_t^{(a)}L_{ab}F^{(b)tr}]\eea

Now we proceed to compute the partition function for the
electrically charged black hole. It is necessary to specify the
time integral of the fourth component of the vector potential on
the boundary. We follow the prcedure of
[8] to carry on this computation for
the problem at hand where the effect of 28 gauge bosons are to
be taken into account. The gauge potential and the electric
field strength have the following form for asymptotically large $r$

\bea \vec A_{tL/R} &=&{\vec\omega_{L/R}\over \beta e}(1-
m{\lambda_{L/R}\over r})\nonumber\\ \vec F_{trL/R}&=&{m
\lambda_{L/R}~\vec\omega_{L/R}\over \beta e}{1\over r^2}\eea

A few comments are in order at this point: the subscript L(R)
refer to the gauge fields arising from the compactification of
the left(right) moving string coordinates. We recall that 22
gauge fields arise from compactification of left hand sector
and other 6 of them come from the right hand sector. Here
$\vec\omega_{L/R}$ are the generalized version of the parameter
introduced in [8]. Here $\beta$ is the inverse
temperature and we have introduced the parameter $e$ to keep
track of some power
countings; however, note that $e$ will not appear in our
expressions for partition function and entropy. The equations
(21) are used in (20) to compute the boundary term
contributions. The constants $\lambda_{L/R}$ have the following
form

\bea \lambda_L={\cosh\gamma\over\cosh\alpha+\cosh\gamma}
 \qquad \lambda_R={\cosh\alpha\over\cosh\alpha+\cosh\gamma}\eea

The geometry is asymptotically flat and the boundary is taken to
be at $r=\infty$. As we have to subtract the flat space
contribution according to the prescription of Gibbons and
Hawking [10,8], we shall first take the radius to be large and finite
and eventually take the limit $r\to\infty$ after the
subtraction.

\bea S_{boundary}-S^{flat}_{boundary}&=&{1\over 2}\beta
(g_{rr})^{-1/2}\partial_r[\Delta^{1/2}\{(g_{rr}^{-1/2}-1\}]
\nonumber\\&=&{1\over 2}\beta(M-m)\eea
where $M$ is the mass of the black hole given by (9). The
contribution of (20) together with (23) can be written as

\bea Z(\beta,\vec\omega)=\exp[-{1\over 2}\beta(M-m)-{1\over 2e^2}
{\lambda_L\vec\omega_L^2+\lambda_R\vec\omega_R^2\over
4\pi(\cosh\alpha +\cosh\gamma)}]\eea

This equation can be interpreted as a grand canonical partition
function for electrically charged black holes where
$\vec\omega$, collectively representing~ $\vec\omega_{L,R}$, play
the role of chemical potential. In order to derive the partition
function with specific charge configurations, we have to
introduce a generalization of the projection technique of
[8].

\bea Z(\beta,\vec Q)&=&e^{-\beta F(\beta,\vec Q)}\nonumber\\&=&
\int_{\infty}^{\infty}\Pi_a {d\omega^{(a)}\over
2\pi}\exp[-{i\over e}(\vec
\omega_L\cdot\vec Q_L+\vec\omega_R\cdot\vec
Q_R)]~Z(\beta,\vec \omega)\eea
where $\vec Q$ stands for both $\vec Q_L$ and $\vec Q_R$. This
integral can be evaluated by the saddle point approximation after
inserting the expression for $Z(\beta,\vec\omega)$ given by
(24). Thus

\bea F(\beta,\vec Q)&=&{1\over 2}\beta(M-m)+2\pi(\cosh\alpha+\cosh
\gamma)\left({\vec Q_L^2\over\beta\lambda_L}+
{\vec Q_R^2\over\beta\lambda_R}\right)\nonumber\\&=&
{1\over 2}(M-m)+{1\over 2}(\vec\Phi_L\cdot\vec
Q_L+\vec\Phi_R \cdot\vec Q_R)\eea
where $\Phi_{L/R}$ are the twenty two/ six components of the
potentials given in
(16). The thermodynamic potential is given by

\bea \Omega(\beta,\vec\Phi)&=&M-T{\cal S}+\vec\Phi\cdot\vec
Q\nonumber\\
&=&F(\beta,\vec Q)-\vec\Phi\cdot\vec Q\eea

Thus the entropy, ${\cal S}$ is given by

\bea {\cal S}&=&\beta(M-F)\nonumber\\
 &=& {A_H\over 4}+{1\over 2}\beta m\eea

We have used the relation (27) in arriving at (28) . Notice that
in both the extremal limits; case
(I) and  case (II), mentioned earlier,
{\it the entropy
vanishes identically}. It is important to note that the
quantum corrections to this entropy are vanishing due to the
non-renomalization theorems.

Now we proceed to calculate the partition function and entropy
of magnetically charged black holes obtained from the
electrically charged ones by S-duality transformation.
In general axion and dilaton are transformed to new
configurations along with the gauge field strengths as given by
(4). The electrically charged black hole corresponds to
$\lambda_1=0$ and $\phi$ and $A_t^{(a)}$ given by (6) and (7).
Our purpose is to implement a special class of duality
transformation under which $\lambda_1'=0$ and $\phi'=-\phi$ and

\bea F_{\mu\nu}^{(a)}\to -\lambda_2(ML)_{ab}F_{\mu\nu}^{(b)}.\eea
This is accomplished by a simple choice of parameters $a=d=0$ and
$b=-c=-1$. Note that under this transformation  $(\vec Q_L,\vec Q_R)\to
(-\vec Q_L,\vec Q_R)$. As the projected partition function for
the electric case is quadratic in $\vec Q_L$ and $\vec Q_R$,
the electromagnetic contribution to the partition function will
remain unchanged under this transformation.
Notice however, the electromagnetic actions corresponding to these two
cases differ by a sign.
Then we can equate the free energy obtained from this partition function
with $M-T{\cal S}$ to obtain the expression for the entropy;
${\cal S}= {A_H\over 4}+{1\over 2}\beta m$, which is the
same for the electric case.

To summerize: our
results derived from the string effective action
show that both the extremal black holes have vanishing entropy
which is conformitive of the work of Hawking, Horowitz and Ross [11].
In the past it has been proposed that extremal black holes might
be identified with elementery particles [12]/ massive states of the
string theory [13]. Computation of partition functions and
entropy for the extremal case
are protected from quantum corrections due to
presence of N=4 supersymmetry. S-duality is beleived to be exact
symmetry of string theory [14], which has far reaching
consequences. Indeed our work indicates another
interesting implication of S-duality.

\noindent{\bf Acknowledgement}

\noindent We are grateful to Ashoke Sen for very illuminating
discussions. One of us (A. G.) would like to acknowledge the
warm hospitality of the Institute of Physics, Bhubaneswar.

\end{document}